\documentclass[12pt]{article}

\usepackage{scicite}

\input epsf

\newenvironment{sciabstract}{%
\begin{quote} \bf}
{\end{quote}}

\newcounter{lastnote}
\newenvironment{scilastnote}{%
\setcounter{lastnote}{\value{enumiv}}%
\addtocounter{lastnote}{+1}%
\begin{list}%
{\arabic{lastnote}.}
{\setlength{\leftmargin}{.22in}}
{\setlength{\labelsep}{.5em}}}
{\end{list}}

\title{Why the cosmological constant is small and positive}

\author
{Paul J. Steinhardt$^{1\ast}$ and Neil Turok$^{2}$\\
\\
\normalsize{$^1$Joseph Henry Laboratories,
Princeton University, Princeton, NJ 08544, USA}\\ 
\normalsize{$^{2}$DAMTP, CMS, Wilberforce Road, Cambridge, CB3 0WA, UK}\\
}

\date{}

\begin{document} 


\baselineskip24pt


\maketitle


\begin{sciabstract}
Within conventional big bang cosmology, it has proven 
to be very difficult to 
understand  why today's cosmological constant is so small. 
In this paper, we show that a cyclic  model of the universe 
can naturally incorporate a dynamical  mechanism 
that automatically relaxes the value of the cosmological constant,
taking account of contributions to the vacuum density at all energy scales.
Because the relaxation time grows exponentially as the vacuum density
decreases, nearly every volume of space spends an overwhelming majority of
the time at the stage when the  cosmological constant is small
and positive, as observed today. 
\end{sciabstract}


\section{Introduction}

One of the greatest challenges in physics today is 
to explain the small positive 
value of 
the cosmological constant or, equivalently, the energy density of the vacuum.  
The observed value, $7\times 10^{-30}$~g/cm$^3$, is 
over one hundred twenty
orders of
magnitude smaller than the Planck density, $10^{93}$~g/cm$^3$,
as the universe emerges from the big bang, yet its value is thought
to be set at that time. Even more puzzling, the vacuum density 
receives a series of contributions from lower 
energy physical effects, including the electroweak and 
quantum chromodynamics (QCD) transitions, that only 
become significant 
at a later stage.
Explaining today's 
tiny value requires a mechanism capable of  canceling 
many very different
contributions with near-perfect precision.  

One long-standing hope had been to find a 
symmetry \cite{mirror} or quantum gravity effect\cite{hawk,worm} 
that forces the vacuum density to be zero. Another hope had 
been to find a relaxation mechanism driving it to zero in the
hot early universe, as the universe expands. 
These hopes have been hard to reconcile with cosmic inflation
and, in any case, have been dashed by recent 
observations indicating 
that the vacuum density is small, positive and 
very nearly 
constant \cite{perl,kirsh}. Now it is apparent that one does not want a complete 
cancellation of the cosmological constant.  And, in order for a relaxation 
mechanism to operate
within the standard inflationary picture, the relaxation time
must at first be much longer than the Hubble time,
so inflation
can take place; then much shorter than the Hubble time
so that nucleosynthesis and structure formation can occur; 
and then, after that, 
much longer than the Hubble time again so that
the vacuum density is
nearly 
constant today, as observed. Despite many attempts,
no 
simple and compelling 
mechanism has been found.
The frustration has been enough to drive many physicists 
to consider anthropic explanations \cite{lindean,weinan}, 
in which one 
assumes that the vacuum density takes on all possible values 
in different regions of space, but that life is only possible 
in one of the rare regions where the vacuum density
is exponentially small.

In this paper, we point out that a cyclic model of  
the type described in
Refs. \cite{cyclic,cyclic2} re-opens the possibility
of solving the cosmological constant problem with a  
natural, monotonic relaxation mechanism. In these 
models,
each cycle consists of a hot big bang followed by a nearly
vacuous period of dark energy domination, ending with a crunch
which initiates the next bang. The duration of
a cycle is typically of the order of a trillion years.  
There is no known limit to the 
number of cycles that have occurred in the past, so the universe today
can plausibly be exponentially older than 
today's Hubble time, 
and still 
form galaxies and stars as observed today. 
Within this cyclic framework, it is reasonable to consider 
mechanisms for relaxing the cosmological constant whose timescale 
is always far greater 
than  
today's Hubble time.  The 
cosmological constant is exponentially smaller than one might
have guessed based on the big bang picture precisely because
the universe is 
exponentially older
than the big bang estimate, so the cosmological constant has had a very long 
time to reduce in value from the Planck scale to the 
miniscule
value observed today.
Furthermore, we will show that it is natural to have mechanisms in which the 
relaxation time increases exponentially as the vacuum density approaches zero
from above,
resulting in a universe in which nearly every volume of space spends an 
exponentially 
longer 
time in a state with small, 
positive cosmological constant than in any other state.
This is in stark contrast to anthropic explanations according to 
which the only regions of 
space ever capable of producing galaxies, stars, planets and life are 
exponentially rare. 

\section{Dynamical Relaxation: A Worked Example}

As a specific example of a dynamical relaxation mechanism, 
we adapt an idea
first discussed 
by Abbott \cite{Abb} in the context of standard big bang cosmology
(see also
Ref.~(\cite{BT})).
In Abbott's model, the vacuum
energy density of a scalar field gradually decays through 
a sequence of exponentially slow quantum tunneling events,
relaxing an initially large 
positive cosmological constant to a 
small value. 
In spite of some appealing
features, Abbott found that the mechanism failed, as we will explain, 
within the context of a big bang universe, 
essentially because the relaxation occurs far too slowly
compared to a Hubble time.  
In this paper, however, we show that the mechanism becomes
viable within the cyclic universe picture.

Abbott's proposal introduces an axion-like 
scalar field 
$\phi$  coupled to the hidden nonabelian 
gauge fields through a pseudoscalar
coupling $(\phi/f) F ^*F$, with $f$ some high energy 
mass scale.
The theory is assumed to
have a classical symmetry 
\begin{equation}
\label{symm}
\phi \rightarrow \phi + {\rm constant},
\end{equation}
which is softly broken at low energies by various effects. 
Integrating out the
gauge fields induces a potential $- M^4 \cos (\phi/f)$, where
$M$ is the scale where the gauge coupling becomes strong. 
(Fields of this type are commonly invoked to suppress CP-violation in the 
strong interactions
\cite{PQ,WCP1,WCP2} and are also ubiquitous in string theory.)

It is natural for $M$ to be very small, as
a consequence of the slow (logarithmic) running of the
coupling in a nonabelian gauge theory. 
For example, in QCD with six flavors,
$\Lambda_{QCD}= M_{Pl} {\rm exp}(-2 \pi/(7 \alpha_{QCD}(M_{Pl}))\sim
100\,$MeV if the coupling strength at the Planck scale 
$\alpha_{QCD}(M_{Pl})\sim 1/50$. (Here
and below, $M_{Pl}=
(8\pi G)^{-{1\over 2}}$).
In
Abbott's model for the hidden axion field, 
$M$ replaces $\Lambda_{QCD}$ and is similarly 
expressed in terms of the relevant coupling to hidden gauge fields. 
For example, if the hidden sector were exactly like QCD,
taking $\alpha(M_{Pl})\sim 1/75$ would give
$M \sim 10^{-3}\,$eV, a viable value for our model.
(Our choices are less extreme than those in Abbott's paper; 
in the 1980's, his goal 
was to obtain a very small vacuum density,
whereas ours is to explain the observed value.)

\begin{figure}
\begin{center}
\epsfxsize=2.5 in \centerline{\epsfbox{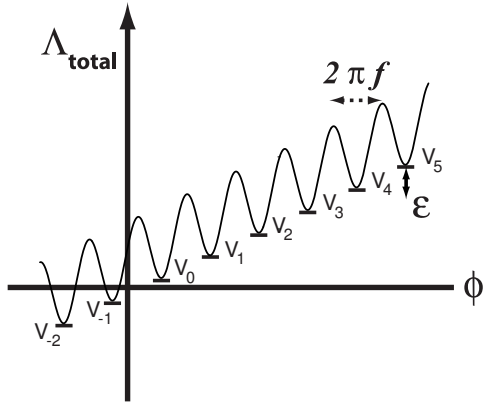}}
  \end{center}
 \caption{
   The effective cosmological constant $\Lambda_{total}$ for
the washboard potential defined in Eq.~\ref{wash} can take discrete
values depending on which minimum $\phi$ occupies.  In the scenario
presented here, the time spent in the lowest positive minimum is 
exponentially greater than the entire time spent in all other minima. 
    }
  \label{washfig}
\end{figure}

The cosine potential 
breaks the symmetry (\ref{symm}) 
down to a discrete subgroup, $\phi \rightarrow \phi
+ 2 \pi N$. The discrete symmetry
is also assumed to be softly broken, 
by a term producing a `washboard' effective potential:
\begin{equation} \label{wash}
V(\phi) = - M^4 \, {\rm cos} \, 
\left( \frac{\phi}{f}\right) + \epsilon \frac{\phi}{2 \pi f} +V_{other},
\end{equation}
where $V_{other}$ includes all other contributions to the 
vacuum density. (The linearity of the second, soft breaking term is
inessential: any potential will do as long as it 
is very gently sloping in the region of interest, around $V=0$.)
Provided $\epsilon <M^4$, (\ref{wash}) has a
set of equally spaced minima $V_N$, with 
effective cosmological constant $\Lambda_{total}$ spaced by
$V_N-V_{N-1}= \epsilon$
(Fig.~\ref{washfig}). No matter what $V_{other}$ is,
there is a minimum with $\Lambda_{total}=V_0$ in the range
 $0 \leq V_0 < \epsilon$. 
Although $\epsilon$ must be chosen to be very small
in order to account for today's tiny vacuum density, this
choice is technically natural within the model since all
quantum corrections to $\epsilon$ are proportional to
$\epsilon$.  Hence, Abbott's model is a 
self-consistent
low-energy effective theory capable of
cancelling contributions to the
vacuum density coming from any other source. 
 
In Abbott's scheme,
the smallness of the cosmological constant today is related through the
relaxation mechanism to the smallness of the parameters 
$M$ and 
$\epsilon$ in the  potential $V(\phi)$.  
Effectively, the intractable problem of 
naturally obtaining an exponentially 
small cosmological constant is transmuted into a tractable problem
of naturally obtaining small axion 
interaction parameters.

Abbott assumed the universe emerges from the big bang 
with some large positive value of $\phi$ and quickly
settles into a minimum with large positive $V_N$,
driving a period of de Sitter expansion 
which dilutes away
any matter and radiation. 
Over time the field $\phi$ then works its way
slowly but inexorably  
downhill. In flat space-time, the tunneling events would occur
at a constant rate, independent of $N$. However, once
the effects of gravity are included, the tunneling rate
becomes slower and slower as $V_N$ decreases. As we shall
see, the universe remains in the
last positive minimum for a relative eternity compared to
the time spent in reaching it. This is the basis for
our claim that the most probable
value for the vacuum density in the model is that 
of the last
positive minimum.

Assuming the field starts high up the potential,
$V_N >> M^2 M_{Pl}^2$, de Sitter 
fluctuations overwhelm the energy barriers
and the field 
makes its way quickly downhill.
But as $V_N$ falls below $M^2 M_{Pl}^2$, 
the barriers become increasingly 
significant and the field progresses downwards
 by quantum tunneling via bubble nucleation
\cite{CD}. Upward tunneling is also allowed,
but hugely suppressed in the parameter
range of interest
\cite{Lee}.

For simplicity, we shall focus on the parameter range
$f^2/M_{Pl}^2 << \beta<1$, where
$\beta 
\equiv \epsilon /M^4$. In the semiclassical
approximation, the rate for nucleating bubbles
of vacuum energy density $V_{N-1}$ beginning from the $V_N$ phase
is 
$\Gamma(N) \propto
{\rm exp}(-B(N))$ where
$B(N)$ is the Euclidean action for the tunneling solution.
In order to describe the scaling of 
$B(N)$ with $N$, we shall neglect unimportant numerical
coefficients and approximate 
$V_N \approx \beta N M^4$.

As $\phi$ tunnels towards minima with decreasing $N$, the nucleation
rate decreases monotonically through three scaling regimes which
match smoothly onto one another:\hfill\break
$\bullet$ For $N
>M_{Pl}^2 /(f^2 \beta)\equiv N_{HM}$,
the de Sitter radius
is smaller than the bubble wall thickness
$\sim f M^{-2}$ and the relevant instanton
is the Hawking-Moss
solution\cite{HM}. In this regime, 
$B(N)\propto
N^{-2}$. \hfill\break
$\bullet$ For $N_{CD} < N < N_{HM}$, where
$N_{CD} \equiv
M_{Pl}^2 \beta /f^2$,
the relevant instantons are of the
Coleman-De Luccia type\cite{CD} and the thin
wall approximation becomes increasingly accurate.
The bubbles are in the scaling regime described
by Parke\cite{Parke}, where the bubble radius is
controlled by gravitational effects. In this
regime,
$B(N) \propto
N^{-{3\over 2}}$. \hfill\break
$\bullet$ As $N$ falls below $N_{CD}$,
the bubble radius becomes much smaller than the
de Sitter radius and the
instantons are well-approximated by
the flat spacetime bubble solution.
Although gravitational effects increase the action 
by only a small factor in this regime,
the correction is very important because $B_0$ is so large.
The leading gravitational correction is given by
\begin{equation}
B(N) = B_0 \left(1- {3\over 2} {(V_N+V_{N-1})T^2 \over M_{Pl}^2 
\epsilon^2}\right), 
\label{beq}
\end{equation}
where the flat spacetime bubble action
$B_0 = {27\over 8} \pi^2 T^4/\epsilon^3$, with
$T$ the wall tension. In a cosine potential this
is $8 M^2 f$.
$B_0$ is an enormous number
$\sim 10^{110}$ for plausible parameters
$f \sim 10^{14}$ GeV, $\beta \sim 0.1$, $M \sim 10^{-3}$ eV.
The gravitational correction causes the 
bubble action to decrease linearly with $N$ in this
final regime.
Thus, as $N$ approaches zero from above, the time
spent at vacuum density $V_N$ scales parametrically
as ${\rm exp} \left(- B_0 (N/N_{CD})\right)$
where $N_{CD}$ is given above. 
For example, with our chosen parameters the time spent at
the last positive value of the vacuum energy
density is more than $10^{10^{110}}$ times
longer than the entire time spent before it.

The whole process ends when the field $\phi$ 
tunnels through to negative potential energy.  Then, the 
negative potential 
causes the space within the bubble to collapse in a time of order 
one Hubble time. 
(For this reason, it makes no difference if the 
field could have tunneled further downhill or not since 
the region will collapse 
before it tunnels further downhill.)  
Space outside the bubble continues to expand from cycle to cycle, so there 
always remain regions with positive cosmological constant. 
Hence, the relaxation process we have described 
naturally leads to a universe which is 
overwhelmingly 
likely to 
possess  
a small positive cosmological constant, in agreement
with observation.
  
Despite its attractive features, the proposal 
proves to be fatally flawed in a 
standard big bang cosmology setting, as Abbott himself pointed 
out, due to the `empty universe problem.'  
Each time the universe is caught in a 
minimum, it undergoes a period of inflation that empties out all matter and 
radiation.  
When a bubble is nucleated, its
interior is nearly empty, too.  At most, it contains an
energy density $\epsilon$ 
and even if this 
is turned entirely into matter and 
radiation it
is far too low to make planets, stars or galaxies.  In fact, whatever 
density does 
lie within the bubble is rapidly diluted away by the next bout
of de Sitter expansion.
The process continues;
new bubbles 
are formed within the old but at each 
stage, the energy density is far 
too small to 
explain the observed universe. 
In effect, the problem is that the relaxation 
process
is too slow for 
standard 
big bang 
cosmology,
so that the universe is empty by the time 
the cosmological constant reaches the requisite value.

\section{Cyclic Model with Dynamical Relaxation}

With this thought in mind, we now turn to 
the cyclic model of the universe
\cite{cyclic,cyclic2}.  According to 
the cyclic picture, the big bang is collision between orbifold planes (branes) 
along an 
extra dimension of space, as might occur in heterotic M-theory\cite{ekpyrotic}. 
A weak, 
spring-like 
force draws the branes together at regular intervals, 
resulting in  
periodic collisions 
that fill the 
universe with new matter and radiation.  After each collision, the branes 
separate and start to re-expand, causing  
the matter and radiation 
to cool and spread out. Eventually, the matter and radiation become so dilute 
that the 
potential energy associated with the inter-brane force takes over.  

In the low 
energy four-dimensional effective theory, the inter-brane distance can be 
described by a 
modulus field $\psi$ which moves back and forth along its effective potential. 
When the 
branes are far apart, the potential energy density is positive and acts as dark 
energy, 
causing the branes to expand at an 
accelerating rate and diluting away the matter 
and 
radiation created at the bang.  At the same time, the 
force draws 
the branes together, causing the potential energy density to decrease from 
positive to 
negative.  As the branes accelerate 
towards one another, their expansion slows.  

Ripples 
in the branes caused by quantum 
fluctuations are amplified by the inter-brane force
as the branes approach one another into a
scale-invariant spectrum of growing
energy density perturbations.  The branes 
remain 
stretched out, though, and any matter and 
radiation within them remains dilute.   
So, after a period of 
a trillion years 
or so, the nearly 
empty  branes 
collide, creating new matter and radiation and initiating 
a new cycle of cosmic evolution. 
In dealing directly with
the big bang singularity, the cyclic scenario poses new challenges
to fundamental theory, and some aspects are still being 
actively debated \cite{deb1,design,deb2,deb3,deb4}. Here, 
we shall assume the cyclic picture is valid.

Now let us suppose we add to this story the axion-like field $\phi$ and 
the associated hidden gauge sector, as entities on one of the two branes. 
Surprisingly, although it
was not invented for this purpose, the cyclic model has just the 
right properties to make 
Abbott's mechanism viable, leading to the prediction we have
emphasized: 
a small, positive cosmological constant.   Four 
features 
inherent to the cyclic model play a key role 
in rendering the
combined model viable:

First and foremost, the cyclic model regularly replenishes the supply of matter 
and 
radiation, instantly solving the `empty universe problem.' Brane
collisions occur every trillion years or so, an infinitesimal time
compared to the eons it takes the universe to tunnel from one minimum 
to 
the next.  
So, between each step down the washboard potential,
the universe undergoes 
exponentially many cycles. 
Each bubble 
that is 
nucleated fills 
with matter and radiation at the cyclic reheat
temperature $T_{reheat} \sim 10^8$~GeV or so \cite{design},
at each new brane collision.
The 
value of $\epsilon$ is 
far
smaller than the energy scale associated withthe collision 
so the 
washboard potential has little effect on reheating. 
Instead, it controls the low energy density, de Sitter like
phase of each cycle, ensuring the cycling solution
is a stable attractor\cite{cyclic2}. 
For $f>> T_{reheat}$, 
$\phi$ is only 
weakly coupled to the matter and radiation, 
and the 
reheating process does not significantly 
affect the evolution of $\phi$. 

A second essential element of the cyclic model is
the orbifold (brane) structure.
If $\phi$ were coupled to the
usual 4d Einstein metric, its 
kinetic energy would be strongly blueshifted during the 
periods of Einstein-frame contraction.  
Instead of proceeding in an orderly manner
down the washboard potential, it would be excited
by the contraction and jump out of the minimum,
accelerating off to infinity as the crunch approached.
In the cyclic model the behavior is quite different because 
$\phi$ 
couples to the induced metric on the brane, 
not the 4d effective Einstein-frame metric.  
The brane expands exponentially from cycle to 
cycle and never contracts to zero;
only the extra dimension that 
separates the branes 
does that.  Consequently, the kinetic energy of $\phi$ is 
red shifted and 
diluted during every cycle, even during the phases when the extra 
dimension (and the Einstein-frame 4d 
effective scale factor) contracts to zero. Thus     
$\phi$ remains trapped in its 
potential
minimum for exponentially long periods until the 
next bubble nucleation occurs.  

The reheating of the universe at the beginning of each cycle
also 
does not excite 
$\phi$ because it is so weakly  coupled. 
In fact, by causing the expansion  to decelerate
and hence suppressing  
the de Sitter fluctuations in $\phi$,  
the matter and radiation actually
decrease the nucleation rate.  The majority of tunneling
events occur during vacuum energy domination, which
is the longest phase of each cycle. 

A third advantageous 
feature of the cyclic model is that, because the homogeneity
and isotropy of the universe and the generation of density perturbations 
are produced by 
very low-energy physics, there is no inflation and, hence, no need to tune the 
relaxation to be slow and then fast.

A fourth critical aspect of the cyclic model
is that dark energy acts as a stabilizer.   By diluting 
the density of matter and radiation and any random excess kinetic 
energy of the branes produced at the previous bounce,
the dark energy ensures that the cycling solution 
is a stable attractor~\cite{cyclic2}. When we add the washboard potential,
the dark energy density 
depends on $V(\phi)$.
The value decreases
by
$\epsilon$ each time a bubble is nucleated.
As long as the dark energy density is positive, the cyclic solution
remains a
stable attractor. Once the sum becomes negative,
the periodic cycling comes to an end.  Most likely,
the interior of the negative potential energy
bubble  collapses
into a black hole, 
detaching itself from  
the universe outside it and ending cycling in that small patch of space;
but the rest of the universe  continues to cycle stably.

Putting these ideas together, the cyclic model 
and Abbott's mechanism are merged 
into a 
new scenario that significantly modifies both.  
In the combined picture, 
there are two fundamental timescales that 
govern the long-term evolution of any patch of universe: the cycling time 
$\tau_{cycle}$ 
and the time it takes to nucleate bubbles, $\tau_N$.  The latter increases 
exponentially as 
the universe tunnels from large $N$ towards $N=0$, and
during each of the stages we have described, $\tau_N$ is
is exponentially greater than 
$\tau_{cycle}$. So, for each jump in $\phi$ the 
universe 
undergoes many cycles and many big bangs.  When $V_N$ is large, the vacuum 
energy 
density dominates the universe at an earlier point in the cycle, before matter 
has a chance 
to cool and form stars, planets or life. But nothing happens to disrupt 
the evolution.  
The universe simply continues cycling 
as $\phi$ continues to hop
down the potential, each step taking exponentially longer than the one before.  
Finally, 
$V_N$ becomes small 
enough that structure begins to form.   How big $N$ is 
before 
this occurs depends on $\epsilon$; for our example above, galaxy 
formation occurs during 
the last few  
hundred  
steps or so.
However, exponentially more time and more cycles are spent at 
$V=V_0$ than 
at any other value.

\section{Discussion}

We have focused on Abbott's particular
mechanism, but we can extract from this
case the conditions that are generally required: (1) a relaxation time
much greater than today's Hubble time; and,
 (2) a dynamics which collapses or recycles
any regions with negative cosmological
constant on a much shorter time scale.  In our example, the relaxation time 
increases as the cosmological constant approaches zero, so that the system
spends most of the time at the lowest positive value. However, it is
also interesting to consider other parameter ranges or other forms 
for $V(\phi)$, including the pure linear  
potential invoked in the 
anthropic model of Ref.~\cite{lindean}, which has no local minimum to be fixed.    
Here the relaxation time {\it decreases} 
as the cosmological constant approaches zero from above.
By introducing cycling and restricting attention to 
the past light cone of any observer, we find that 
most galaxies are produced when the vacuum density is smaller,
but not much smaller than the matter density.

In either example, our result is a universe in which the cosmological 
constant $\Lambda(t)$ is an ultra-slowly varying function of time $t$
and in which 
virtually every patch of space proceeds through stages of evolution that 
include ones in which $\Lambda(t)$ is  small enough to 
be habitable for life.  It is interesting to contrast this 
situation with the anthropic picture, especially versions based on inflationary 
cosmology, 
for which the fraction of habitable space is infinitesimally small.  All other 
things being equal, 
a theory that predicts that life can exist almost everywhere is overwhelmingly 
preferred 
by Bayesian analysis (or common sense) over a theory that predicts it can exists 
almost 
nowhere. 

Although the relaxation time scale is far too slow to be detectable, 
the general picture we have suggested here can be falsified.  First, since
it relies on the cyclic model, it inherits the cyclic prediction for  primordial 
gravitational waves \cite{gwave}.  Second, one might for other
implications of having an exponentially long time for fields or 
couplings to evolve.  For example, axions in QCD and string theory with 
$f \gg 10^{12}$~GeV are well-motivated theoretically, but ruled out in 
conventional inflationary theory because de Sitter fluctuations 
typically excite 
the field to a value where its energy density overdominates the universe 
today \cite{thomas}.  Some propose resolving this dilemma, also, using the 
anthropic principle 
\cite{MTurn,Linde},  but, then, the same reasoning suggests that axions should 
contribute all or most of the dark matter density today \cite{Wil}.  In the 
alternative picture we have presented, though, there is no inflation and 
axions are never excited.  So, finding axions with large $f$ and negligible 
density would be an embarrassment for the inflationary picture but 
would fit naturally in the picture outlined here. 
Similar considerations apply to other solutions to
the strong CP problem \cite{irr} where a very long relaxation time may be 
useful.

\begin{scilastnote}
\item We thank Amol Upadhye, E.J. Weinberg and the participants
of the workshop `Expectations of an Ultimate Theory' held in Trinity
College, Cambridge, in September 2005, for useful discussions. 
This work is supported in part by PPARC (NT) and by  
US Department of Energy grant DE-FG02-91ER40671 (PJS).
\end{scilastnote}

\end{document}